**Modeling COVID-19 Transmission using IDSIM, an Epidemiological-Modelling Desktop App with Multi-Level Immunization Capabilities**

Eleodor Nichita [1,a], Mary-Anne Pietrusiak [2,b], Fangli Xie [3,b], Peter Schwanke [4,a] and Anjali Pandya [5,b]

[a] *Ontario Tech University, 2000 Simcoe Street North, Oshawa, ON, Canada*

[b] *Durham Region Health Department, 605 Rossland Road East, Whitby, ON, Canada*

Abstract

*The COVID-19 pandemic has placed unprecedented demands on local public health units in Ontario, Canada, one of which was the need for in-house epidemiological-modelling capabilities. To address this need, Ontario Tech University and the Durham Region Health Department developed a native Windows desktop app that performs epidemiological modelling of infectious diseases. The app is an implementation of a multi-stratified compartmental epidemiological model that can accommodate multiple virus variants and levels of vaccination, as well as public health measures such as physical distancing, contact tracing followed by quarantine, and testing followed by isolation. This article presents the epidemiological model and epidemiological-simulation results obtained using the developed app. The simulations investigate the effects of different factors on COVID-19 transmission in Durham Region, including vaccination coverage, vaccine effectiveness, waning of vaccine-induced immunity, advent of the Omicron variant and effect of COVID-19 booster vaccines in reducing the number of infections and severe cases. Results indicate that, for the Delta variant, natural immunity, in addition to vaccination-induced immunity, is necessary to achieve herd immunity and that waning of vaccine-induced immunity lengthens the time necessary to reach herd immunity. In the absence of additional public health measures, a wave driven by the Omicron variant is predicted to pose significant public health challenges with infections predicted to peak in approximately two to three months, depending on the rate of administration of booster doses.*

## 1. INTRODUCTION

In the summer of 2020, as the province of Ontario, Canada, was recovering from the first wave of COVID-19 infections, it became clear that a second wave was developing. Local public health units were called upon to make forecasts about the future evolution of cases and recommend public health interventions at a time when modelling resources, both computational and human, were scarce. Modelling results were usually prepared at the national or provincial level by sizeable teams of epidemiologists and mathematicians with results being only partially applicable to local situations. The Regional Municipality of Durham, which comprises areas to the east of Toronto and has a population of approximately 750,000, was facing challenges common to all Ontario public health units (PHUs). To alleviate the shortage of modelling resources, Durham Region Health Department established a collaboration with Ontario Tech University to develop in-house COVID-19 epidemiological modelling capabilities. The immediate objective was to create a model and software package in the form of a Windows desktop app to be used by staff epidemiologists for making forecasts and informing policy decisions without the need for high-performance computing systems or extensive training.

[1] Eleodor.Nichita@ontariotechu.ca
[2] Mary-Anne.Pietrusiak@durham.ca
[3] Fangli.Xie@durham.ca
[4] Peter.Schwanke@ontariotechu.net
[5] Anjali.Pandya@durham.ca

For simplicity and practicality, a dynamic compartmental (deterministic) model developed by the Public Health Agency of Canada (Ogden, et al., 2020) was initially adopted. This initial model consisted of seven compartments (susceptible, exposed, exposed quarantined, infectious, infectious isolated, hospitalized, and removed), and allowed for only one ancestral strain. It was implemented as a *Modern Fortran* (Fortran with object-oriented programming features) code with an Excel/Visual-Basic user interface. As variants emerged and vaccines became available, additional capabilities were added to the model and the implementation was switched to a native MS-Windows desktop app with a *Modern Fortran* computational backend. The app was named IDSIM (**I**nfectious-**D**isease **SIM**ulator). This work presents the current (November 2021) IDSIM model and illustrates some of its capabilities by performing four simulations of COVID-19 transmission under different conditions.

## 2. EPIDEMIOLOGICAL MODEL

The epidemiological model is a multi-stratified compartmental model that can accommodate multiple virus variants and levels of vaccination, as well as public health measures such as physical distancing, contact tracing followed by quarantine, and testing followed by isolation.

### 2.1 Compartments and flowchart

The diagram of the epidemiological model is shown in Fig. 1.

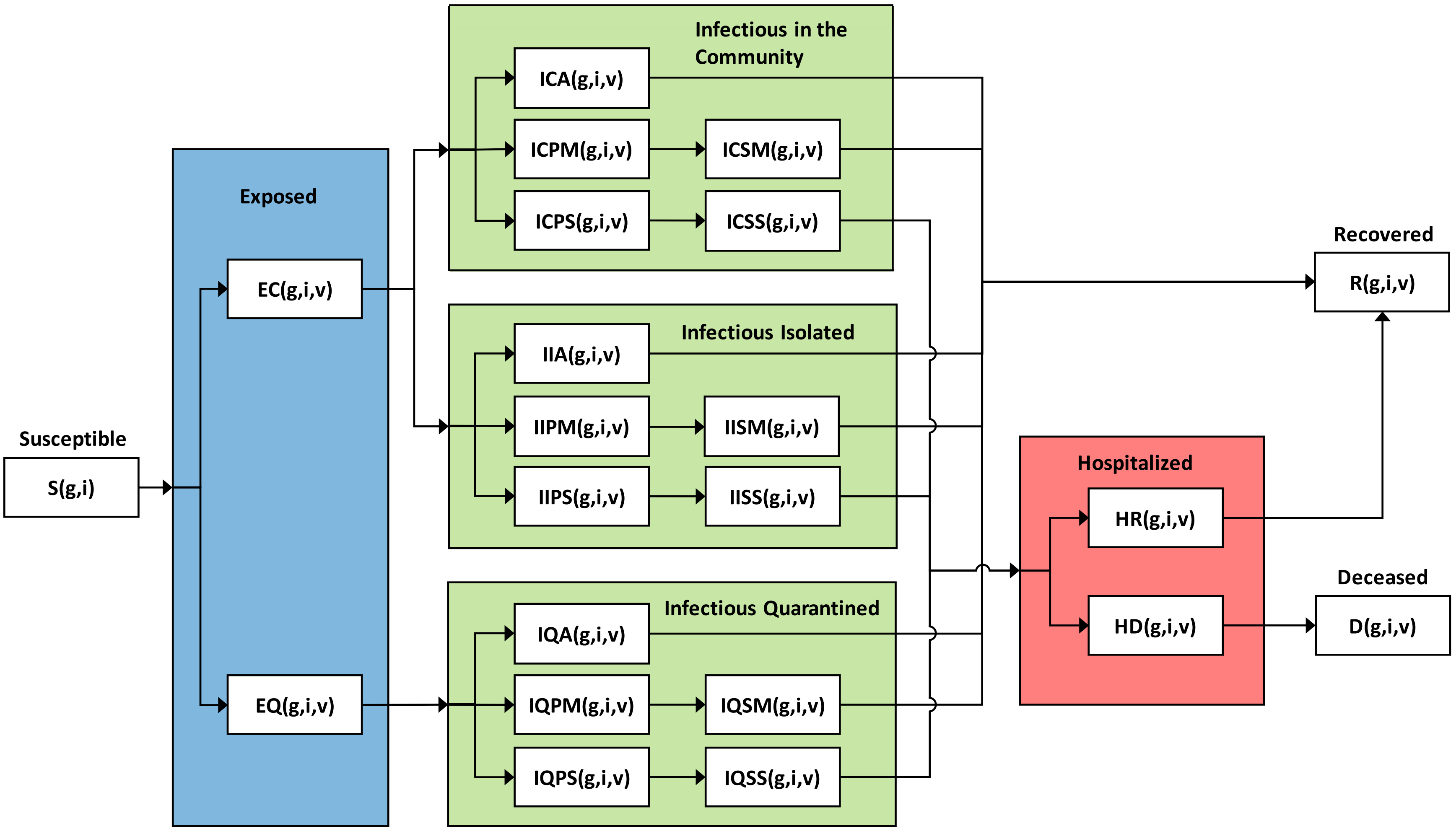

Figure 1: Compartmental Model Diagram

The compartment notations are as follows:

1. S = **S**usceptible

   **E**xposed

2. EC = **E**xposed in the **C**ommunity (not quarantined)
3. EQ = **E**xposed **Q**uarantined

   **I**nfectious

   **I**nfectious in the **C**ommunity (not isolated and not quarantined)

4. ICA = **I**nfectious, in the **C**ommunity, **A**symptomatic
5. ICPM = **I**nfectious, in the **C**ommunity, **P**re-symptomatic, will progress to **M**ild symptoms
6. ICPS = **I**nfectious, in the **C**ommunity, **P**re-symptomatic, will progress to **S**evere symptoms
7. ICSM = **I**nfectious, in the **C**ommunity, **S**ymptomatic, **M**ild symptoms
8. ICSS = **I**nfectious, in the **C**ommunity, **S**ymptomatic, **S**evere symptoms

   **I**nfectious **I**solated

9. IIA = **I**nfectious **I**solated **A**symptomatic
10. IIPM = **I**nfectious **I**solated **P**re-symptomatic, will progress to **M**ild symptoms
11. IIPS = **I**nfectious **I**solated **P**re-symptomatic, will progress to **S**evere symptoms
12. IISM = **I**nfectious **I**solated **S**ymptomatic **M**ild
13. IISS = **I**nfectious **I**solated **S**ymptomatic **S**evere

   **I**nfectious **Q**uarantined

14. IQA = **I**nfectious **Q**uarantined **A**symptomatic
15. IQPM = **I**nfectious **Q**uarantined **P**re-symptomatic, will progress to **M**ild symptoms
16. IQPS = **I**nfectious **Q**uarantined **P**re-symptomatic, will progress to **S**evere symptoms
17. IQSM = infectious quarantined symptomatic mild
18. IQSS = infectious quarantined symptomatic severe

**H**ospitalized

19. HR = **H**ospitalized **R**ecovering

20. HD = **H**ospitalized **D**ying

21. R = **R**ecovered

22. D = **D**eceased

The population in each compartment is categorized by *combined stratum* (subscript *g*), immunization status (subscript *i*) and variant (subscript v). The differential equations governing transition from one compartment to another are presented in the Appendix.

### *Variants*

The variant subscript, v, applies to all compartments other than the one comprised of susceptible individuals.

### *Combined strata*

Each *combined stratum* is a combination of strata corresponding to multiple stratifications. For example, if a particular population were stratified by age into two strata, those under 50 and those 50+, and by gender into two strata, female and male, then subscript *g* would take values between 1 and 4, corresponding to the four *combined strata*: females under 50, females 50+, males under 50, and males 50+.

### *Immunization status*

The immunization status can have as many levels as necessary, identified by subscript *i*. For example, subscript *i* could take values between 1 and 5, with the following meanings:

1. not vaccinated
2. first dose administered, first-dose protection not yet achieved
3. first-dose protection achieved
4. second dose administered, second-dose protection not yet achieved
5. second-dose protection achieved

Persons advance from one immunization level to the next either through vaccination or the passing of time. Using the example above, individuals would move from level 1 to level 2 and from level 3 to level 4 through vaccination (defined by the number of people being vaccinated daily), and from level 2 to level 3 and from level 4 to level 5 by the simple passing of time (defined by the average time necessary to achieve protection after vaccination). This is illustrated in Fig. 2.

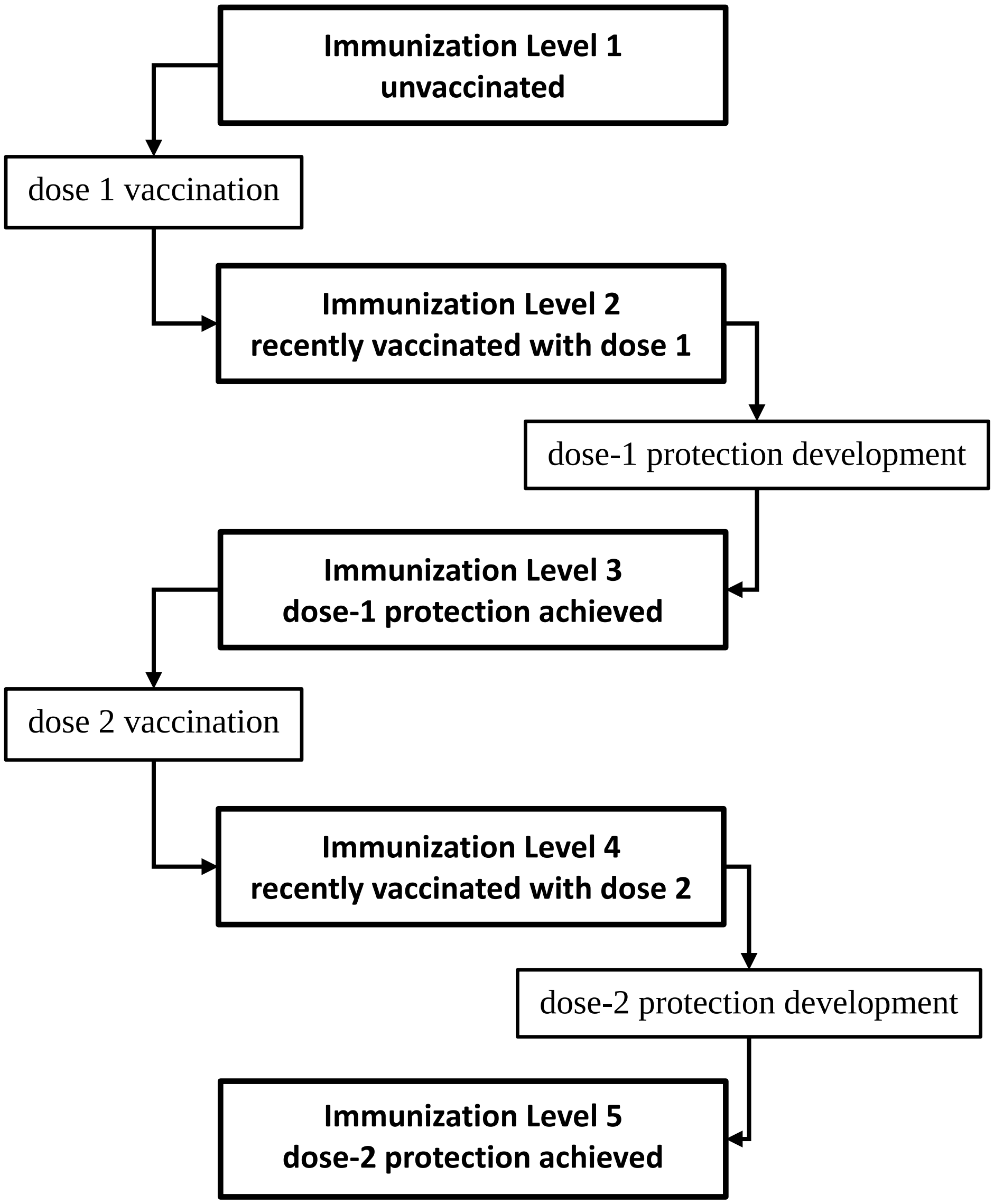

Figure 2: Example of Immunization Levels

## 2.2 Transmission model

Persons become exposed through contact with one or multiple infectious individuals. The exposure rate is characterized by the number of contacts per individual per day and by the probability of transmission with contact. The former is characterized by an average number of daily contacts for the population. The latter is characterized by an average number for the population which is then modulated by variant-dependent and vaccination-level-dependent factors.

### 2.3 Stratification parameters

Each stratification can have a different number of strata. Each stratum is defined by the following parameters:

1. Fraction of population belonging to the stratum
2. Susceptibility modulator (a factor that multiplies the probability of transmission with contact for susceptible individuals belonging to the stratum).
3. Severity modulator (a factor that multiplies the fraction of symptomatic individuals in the stratum that go on to develop severe symptoms). For example, in the 80+ age group, a value greater than 1 would be appropriate to represent the higher probability of severe outcomes for that age group.

### 2.4 Variant parameters

Each variant, including the ancestral strain, is defined by the following parameters:

1. Latency time (since exposure)
2. Incubation time (since exposure)
3. Time to hospitalization for severe cases (since exposure)
4. Time to recovery for non-severe cases (since exposure)
5. Time to recovery after hospitalization (for severe cases that recover)
6. Time to death after hospitalization (for severe cases that do not recover)
7. Probability of transmission with contact
8. Fraction of infectious individuals that are symptomatic
9. Fraction of infectious symptomatic individuals that have severe symptoms
10. Fraction of hospitalized individuals that recover

### 2.5 Immunization-level parameters

Each immunization level is defined by the following parameters:

1. Transmissibility factor (a factor, usually less than or equal to 1, that multiplies the probability of transmission with contact for infectious individuals with a specific vaccination level)

   > For infectious individuals who are unvaccinated or recently vaccinated (before developing protection), this factor would be 1. For individuals who are both vaccinated and infectious and who have already developed some protection, the factor would normally be less than 1 to represent the fact that those individuals are less contagious.

2. Susceptibility factor (a factor, usually less than or equal to 1, that multiplies the probability of transmission with contact for susceptible individuals with a specific vaccination level)

For susceptible individuals who are unvaccinated or recently vaccinated (before developing protection), this factor would be 1. For susceptible individuals who have already developed some protection, the factor would normally be less than 1 to represent the fact that those individuals are less likely to become infected. This factor is essentially equal to one minus the vaccine efficacy.

3. Severity factor (a factor, usually less than or equal to 1, that multiplies the fraction of symptomatic individuals with severe symptoms)

   For infectious symptomatic individuals who are unvaccinated or recently vaccinated (before developing protection), this factor would be 1. For infectious symptomatic individuals who have already developed some protection, the factor would normally be less than 1, to represent the fact that those individuals are less likely to develop severe symptoms.

4. The rate at which individuals move from one immunization level to the next, expressed as either:
   a. Persons vaccinated per unit time (day)
   b. Average time (days) before protection changes following vaccination

### 2.6 Parameters for public health measures

Public health measures are characterized by the following parameters:

1. Fraction of exposed individuals that are successfully quarantined
2. Fraction of infectious individuals that are tested and successfully isolated
3. Coefficient for additional unspecified public health measures. This general factor, usually less than or equal to 1, appears in the force of infection to account for measures such as mask wearing or physical distancing. It can also be manually adjusted to fit model predictions to actual recorded data.

### 2.7 Modelling of decrease in vaccine protection over time and of third dose

The multi-level immunization status can be used to model the decrease in vaccine protection and subsequent need for third doses once the protection has decreased to a certain level. An example is to use eight immunization levels as follows:

1. not vaccinated
2. first dose administered, protection after first dose not yet achieved
3. first dose administered, protection after first dose achieved
4. second dose administered, protection after second dose not yet achieved
5. second dose administered, protection after second dose achieved
6. second-dose protection decreased
7. third dose administered, protection after third dose not yet achieved

8. third dose administered, protection after third dose achieved

Progression from level 5 to level 6 happens through the passage of time (e.g., three months for a 20% decrease in vaccine protection). Progression from level 6 to level 7 happens through administration of the third vaccine dose. Progression from level 7 to level 8 happens through the passage of time (e.g., two weeks for increased protection to develop).

### 2.8 Model assumptions

In its current form, the model makes several assumptions:

1. Recovery from one variant offers full and permanent immunity against all variants.
2. Breakdown by strata in a particular stratification is independent of the other stratifications. As with the previous stratification example, if 50% of the population were female and 50% of the population were male, that is assumed to be true both for persons under 50 and for persons 50+. Similarly, if 60% of the population is under 50 and 40% is 50+, then that is assumed to be true for both male and female populations.
3. All severe cases are hospitalized.
4. The number of contacts per day per person is the same for all combined strata and independent of the vaccination level of an individual. Quarantined, isolated and hospitalized individuals are assumed to have no contacts.

### 2.9 Simulation starting and end points

Initial conditions at "Day 0" can be specified in detail, including the population of each compartment by stratum, vaccination status and variant. This allows simulations to start from realistic data acquired in the field rather than from generic assumptions of one infectious individual. The end point of a simulation can be saved and used as the starting point of a new simulation, thus allowing the indefinite extension of the simulation time interval.

### 2.10 Time-dependent epidemiological parameters

Time-dependent parameters can be simulated by assuming them to be constant over finite time intervals, with step changes from one interval to the next. For example, the simulation of an entire year can be performed in 30-day intervals, with parameters updated at the start of each such simulation interval.

## 3. COVID-19 TRANSMISSION SIMULATIONS

Four simulations are performed, to investigate the effect of specific factors on COVID-19 transmission in Durham Region:

- Simulation 1: Effect of different vaccination coverage values with vaccination being the only public health measure, assuming no waning of vaccine-induced immunity over time
- Simulation 2: Effect of different vaccine effectiveness values with specified public health measures in place, assuming no waning of vaccine-induced immunity over time
- Simulation 3: Effect of waning vaccine-induced immunity after three months, assuming specified public health measures
- Simulation 4: Effect of the advent the Omicron variant, and impact of COVID-19 booster vaccines on transmission and severity of Delta and Omicron variants assuming specified health measures and waning of vaccine-induced immunity over time

Pfizer-BioNTech (Comirnaty, BNT162b2) and Moderna (Spikevax, mRNA-1273) are the two main types of COVID-19 vaccines offered in Durham Region. Vaccine effectiveness against COVID-19 infection drops between three and six months. It is assumed that individuals in Durham Region received their second dose, on average, four months prior to the simulation start date.

General simulation parameters are shown in Table 1. Parameters specific to individual simulations are shown in Table 2. Day 0 of the simulation is always November 22, 2021.

Table 1: General Simulation Parameters

| **Parameter** | **Value** | **Source** |
|---|---|---|
| Number of strata | 1 | user specified |
| Number of infectious individuals on Day 0 | 200 active cases reported on Day 0. For each reported case, there are three undetected cases in the community, for a total of 800 infectious individuals. | PHU data,<br>(Public Health Ontario, 2020) |
| Number of contacts per day | 10 | (Mossong, et al., 2008)<br>(Feehan & Mahmud, 2021) |
| Latency period | 3 days for Delta<br>1 day for Omicron | (Li, et al., 2021)<br>(Jansen, et al., December 2021)<br>(Hay, et al., 2022) |
| Incubation time | 5 days for Delta<br>3 days for Omicron | (Wang, et al., 2021) (Li, et al., 2021)<br>(Jansen, et al., December 2021)<br>(Hay, et al., 2022) |

| Parameter | Value | Source |
|---|---|---|
| Time to hospitalization (from exposure) | 10 days for Delta & Omicron | PHU data |
| Time to recovery for non-severe (from exposure) | 14 days for Delta & Omicron | PHU data |
| Time to recovery after hospitalization | 14 days for Delta<br>10 days for Omicron | PHU data<br>(Hay, et al., 2022) |
| Time to death after hospitalization | 15 days for Delta & Omicron | PHU data |
| Probability of transmission with contact | 0.058 for Delta (estimated based on $R_0$, infectious period and contact rate)<br>0.232 for Omicron | (Liu & Rocklöv, 2021)<br>(Ito, et al., 2021) |
| Fraction symptomatic (of infectious) | 0.85 for Delta & Omicron | PHU data |
| Fraction severe (of symptomatic) | 0.03 for Delta<br>0.012 for Omicron | PHU data<br>(Public Health Ontario, 2021) |
| Fraction recovered after hospitalization | 0.6 for Delta<br>0.8 for Omicron | PHU data<br>(Public Health Ontario, 2021) |
| Transmissibility factor for unvaccinated | 1 for Delta & Omicron | (Eyre, et al., 2021) |
| Susceptibility factor for unvaccinated | 1 for Delta & Omicron | (Eyre, et al., 2021) |
| Severity factor for unvaccinated | 1 for Delta & Omicron | (Public Health Ontario, 2021) |
| Transmissibility factor after 1 dose | 0.8 for Delta & Omicron | (Eyre, et al., 2021) |
| Susceptibility factor after 1 dose | 0.7 for Delta & Omicron | (Eyre, et al., 2021)<br>(Nasreen, et al., n.d.) |
| Severity factor after 1 dose | 0.3 for Delta & Omicron | (Public Health Ontario, 2021) |
| Transmissibility factor after 2 doses | 0.5 for Delta<br>0.6 for Omicron | (Eyre, et al., 2021) |
| Susceptibility factor after 2 doses | 0.2 for Delta<br>0.6 for Omicron | (Eyre, et al., 2021)<br>(UK Health Security Agency, 2021)<br>(Andrews, et al., 2021) |
| Severity factor after 2 doses | 0.2 for Delta<br>0.3 for Omicron | (Thompson, et al., 2022)<br>(Jansen, et al., December 2021) |
| Transmissibility factor after 3 doses | 0.5 for Delta<br>0.5 for Omicron | (Eyre, et al., 2021) |

| Parameter | Value | Source |
|---|---|---|
| Susceptibility factor after 3 doses | 0.1 for Delta<br>0.3 for Omicron | (UK Health Security Agency, 2021) |
| Severity factor after 3 doses | 0.06 for Delta<br>0.1 for Omicron | (Thompson, et al., 2022)<br>(Jansen, et al., December 2021) |
| Fraction of population with 1 dose on Day 0 | 0.01-0.03 | PHU data |
| Fraction of population with 2 doses on Day 0 | 0.72-0.74 | user specified |
| Fraction of population with 3 doses on Day 0 | 0 | user specified |
| Number of exposed individuals on Day 0 | 218 | PHU data |
| Infectious period | 11 days for Delta<br>13 days for Omicron | calculated based on recovery period |
| Population | 738,000 | census data |
| Number of recovered persons on Day 0 | 110700 | PHU data |
| Number of deceased persons on Day 0. | 389 | PHU data |

Table 2: Specific Simulation Parameters

| | **Simulation 1: Different vaccination coverage values** | **Simulation 2: Different vaccine effectiveness values** | **Simulation 3: Cases with and without waning immunity** | **Simulation 4: Evolution of Delta and Omicron variants and effect of booster doses** |
|---|---|---|---|---|
| Distribution of Variants on Day 0 | 100% Delta | | | 99% Delta and 1% Omicron |
| Vaccination levels | 1) Unvaccinated<br>2) Two doses | 1) Unvaccinated<br>2) Dose 1<br>3) Dose 2 | 1) Unvaccinated<br>2) Dose 1<br>3) Dose 2<br>4) Reduced immunity | 1) Unvaccinated<br>2) Dose 1<br>3) Dose 2 (reduced immunity)<br>4) Dose 3 |

| | **Simulation 1: Different vaccination coverage values** | **Simulation 2: Different vaccine effectiveness values** | **Simulation 3: Cases with and without waning immunity** | **Simulation 4: Evolution of Delta and Omicron variants and effect of booster doses** |
|---|---|---|---|---|
| Vaccination coverage | Compare transmission under different vaccination proportions:<br>• 70%,<br>• 80%,<br>• 90% and<br>• 100% | • Start with a vaccination coverage of 2% for Dose 1, 72% for Dose 2<br>• 400 Dose 1 administrated per day, and<br>• 300 Dose 2 administrated per day | | • Start with a vaccination coverage of 1% for Dose 1, 74% for Dose 2 and 0% for Dose 3<br>• 500 Dose 1 administrated per day,<br>• 200 Dose 2 administrated per day, and<br>• 3000 or 5000 Dose 3 administered per day |
| Public Health Measures | Public health measure co-efficient is 1, no isolation and no quarantine | • Public health measure coefficient is 0.8 (fitted to match the estimated value of Rt ≈1 for Durham Region on Day 0)<br>• All the cases reported are isolated and<br>• 5% of exposed are quarantined | | |
| Vaccine Effectiveness | Vaccine effectiveness is 80%,<br>Vaccine reduces transmission by 50% after Dose 2 and reduces severity by 85% | Compare the following:<br>• Effectiveness rate of 33% for Dose 1, and 80% for Dose 2<br>• Effectiveness rate of 56% for Dose 1, and 87% for Dose 2 | Effectiveness at two weeks: 33% for Dose 1, 80% for Dose 2;<br>Transmissibility at two weeks: 83% for Dose 1, 50% for Dose 2; | • Omicron variant is four times as transmissible as the Delta variant |
| Waning immunity after vaccination | None | None | Vaccine Effectiveness 12 weeks Dose 2: 45%,<br>Transmissibility 12 weeks after Dose 2: 76% | Reduced protection for individuals who had only two doses of vaccine |

## 4. RESULTS AND DISCUSSION

The first simulation quantifies the impact of different vaccination proportions on COVID-19 transmission when vaccination is the only public health control measure and there are no additional vaccinations during the simulation period.

Figure 3 shows daily reported new infections for different vaccination proportions. Results show that the number of daily reported COVID-19 cases significantly decreases with increased vaccination proportion. However, even with an 80% vaccination coverage, there is still a very high daily number of reported cases. At least 90% of the population needs to be vaccinated to control an epidemic consisting of the Delta variant. In reality, it is hard to reach such high vaccination coverage. The results indicate that even small increases in vaccination coverage will significantly reduce COVID-19 transmission but that other control measures are also needed. Public health control measures other than vaccination include case detection, contact tracing and quarantine, physical distancing, limiting social gatherings, mask use, self-monitoring and other "lockdown" measures.

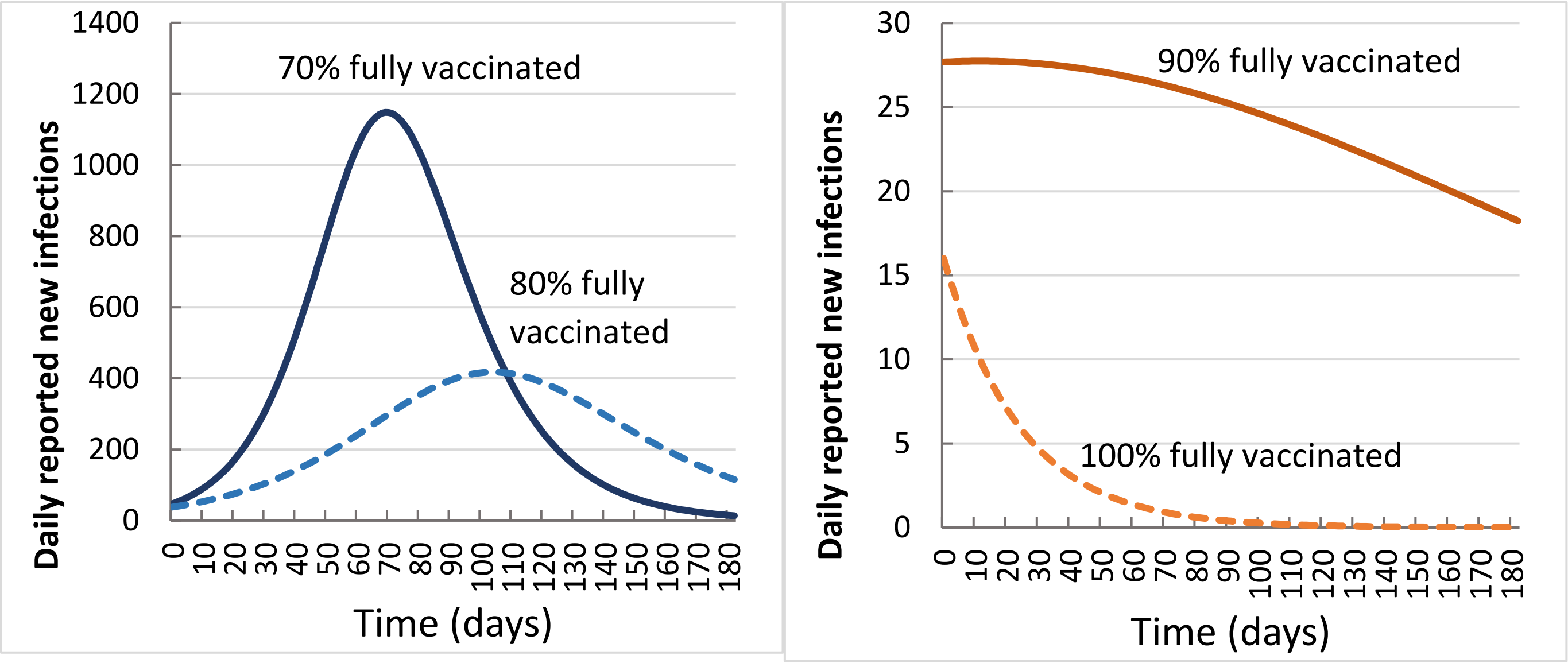

Figure 3: Impact of different vaccination coverage values on COVID-19 transmission

The second simulation evaluates the impact of two different vaccine effectiveness values on COVID-19 transmission under the specific public health control measures in effect in Durham Region in November 2021, as shown in Table 2. The two sets of values for vaccine effectiveness come from a UK study (Eyre, et al., 2021) and an Ontario study (Nasreen, et al., n.d.). The Ontario study found a higher vaccine effectiveness than the UK study. Under the control measures in effect in November, the effective reproduction number, Rt, estimated based on daily case data, was approximately 1.0. The public health measure coefficient is manually fitted so, on Day 0, the predicted Rt matches the estimated Rt $\approx$ 1 in Durham Region for the month of November 2021. The population vaccination fraction starts with 2% for dose 1 and 72% for dose 2. Each day, 400 people are vaccinated with the first dose and 300 people are vaccinated with the second dose. By the end of the 180-day simulation period, 92% of the population is fully vaccinated. Simulation results for the two sets of vaccine effectiveness data are shown in Fig. 4.

With current (November 2021) vaccination and other public health control measures, it is projected that the daily new infections (assuming only Delta variant) will decrease over time. However, at day 90, the projection based on the UK vaccine-effectiveness data shows twice the number of daily reported infections than the projection based on the Ontario vaccine-effectiveness data.

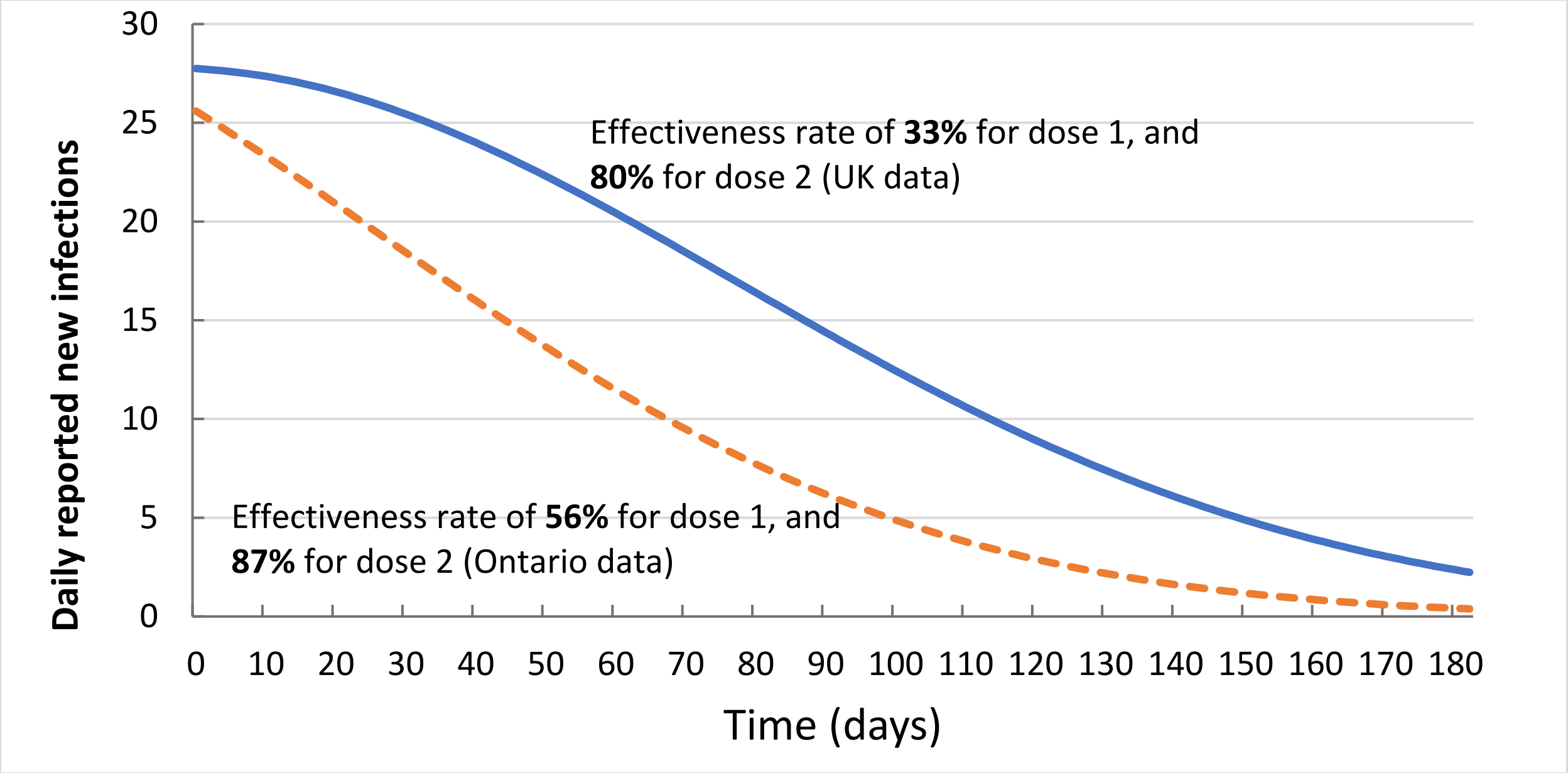

Figure 4: Impact of Different Vaccine Effectiveness Values on COVID-19 Transmission

In addition to comparing the effectiveness of different control measures, the modelling application can also be used to understand the impact of waning vaccine-induced immunity on COVID-19 transmission.

The third simulation estimates the impact of decreasing vaccine effectiveness over time. It compares the case of no immunity waning to the case of immunity waning after three months. Results are shown in Fig. 5.

Under the current (November 2021) vaccination program and other public health measures, if there is no waning of immunity after vaccination, the epidemiological curve is projected to be flattened from the beginning and daily reported new infections to be decreasing. With waning immunity, the likelihood of being protected from COVID-19 infection (vaccination effectiveness) decreases and the likelihood of fully-vaccinated people transmitting the disease increases. If there is waning immunity, the number of daily reported new infections is higher. The epidemiological curve flattens six months into the simulation period.

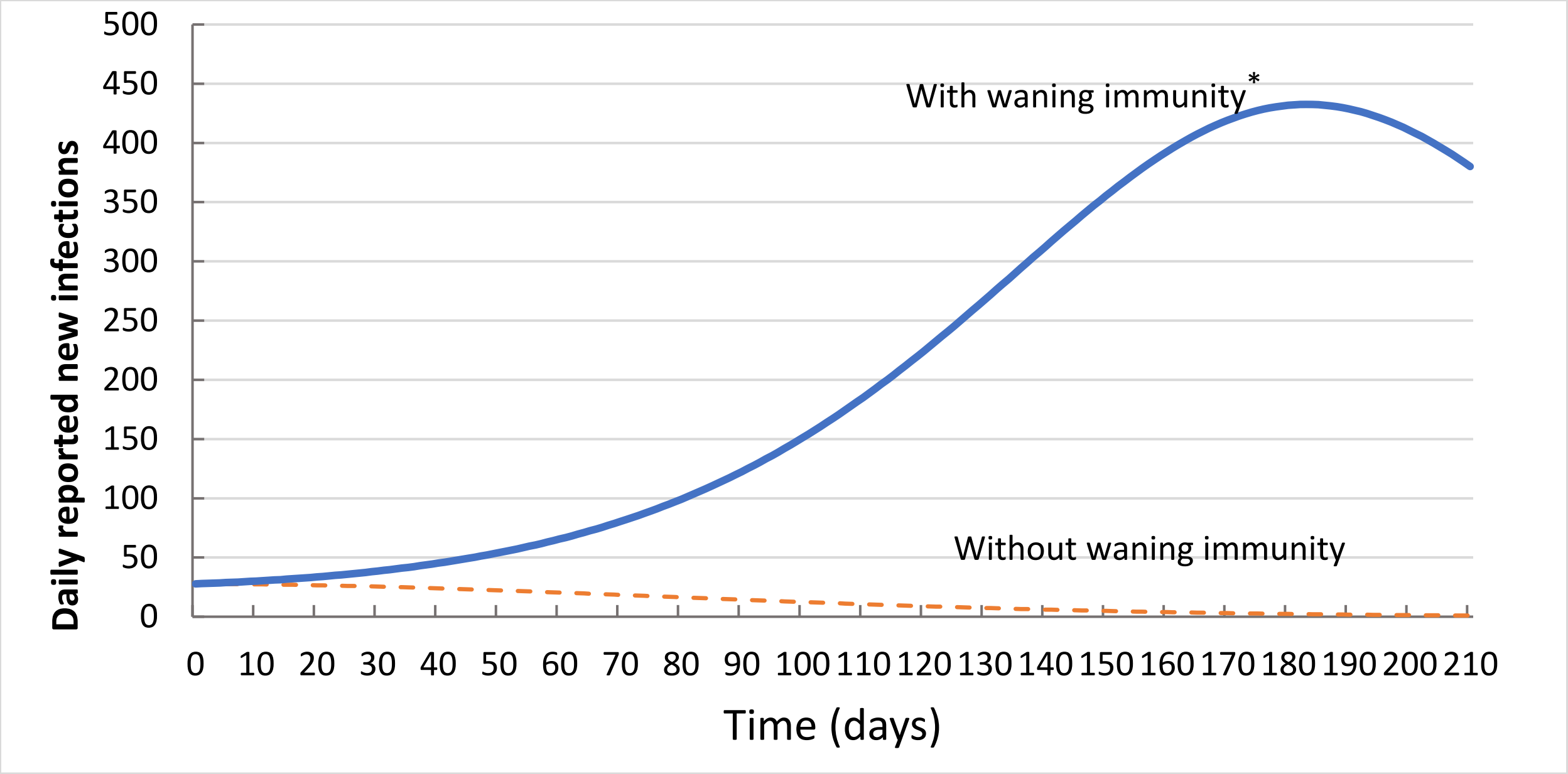

* 12 weeks after full vaccination, vaccine effectiveness is assumed to decrease from 80% to 45% and the reduction of transmissibility in vaccinated people is assumed to drop from 50% to 24%

Figure 5: Impact of waning vaccine-induced immunity on COVID-19 transmission

The modelling application can also simulate disease transmission with multiple variants. The fourth simulation investigates the advent of the Omicron variant in addition to the Delta variant, as well as the effectiveness of a third dose (booster) of mRNA vaccine in preventing COVID-19 infection, hospitalization and death. It is assumed that the Omicron variant is four times as transmissible as the Delta variant (Ito, K 2021; Nishiura H 2022). On Day 0 of the simulation, 99% of the existing infections are assumed to be due to the Delta variant, and 1% due to the Omicron variant. COVID-19 infections, hospitalizations, and deaths are compared for three scenarios: 1) no third (booster) dose of mRNA vaccine, 2) 3000 booster doses administrated per day and 3) 5000 booster doses administrated per day. It is assumed that the booster-dose coverage starts at 0% on Day 0 of the simulation and booster doses are administered until booster coverage reaches 93% for the eligible population (18 years of age or older). It takes 180 days to reach 93% booster coverage if 3000 booster doses are administered per day and 110 days to reach that coverage level if 5000 booster doses are administered per day.

Figure 6 shows the impact of the Omicron variant and the third dose of vaccine on disease transmission. The number of new Omicron-variant infections is projected to surpass the number of new Delta-variant infections after just two weeks from Day 0, in the middle of December 2021. Within a month, Omicron is projected to become the dominant variant and account for the majority (97%) of infections. Similar results have been found by the Ontario COVID-19 Science Advisory Table (Ontario COVID-19 Science Advisory Table, 2022).

Simulation results show that booster doses have a dramatic impact on COVID-19 related infections, hospitalizations (including in-patients and ICUs) and deaths (Figures 6, 7, and 8).

Over a quarter of infections (26%) are prevented if 3000 booster doses are administered each day in Durham Region, and 41% of infections are prevented if 5000 booster doses are administered each day. Administering 5000 booster doses each day also prevents more than half of the hospitalizations and almost half of the deaths (Table 3, Table 4 and Figure 8).

Figure 8 shows the number of daily reported new infections, patients in hospital on a given day, and total deaths by vaccination status for the 5000-booster-dose-per-day scenario. Although vaccinated people are predicted to account for almost three-quarters of the COVID-19 infections by day 180, they are predicted to account for only 30% of severe cases (measured by hospitalizations and deaths).

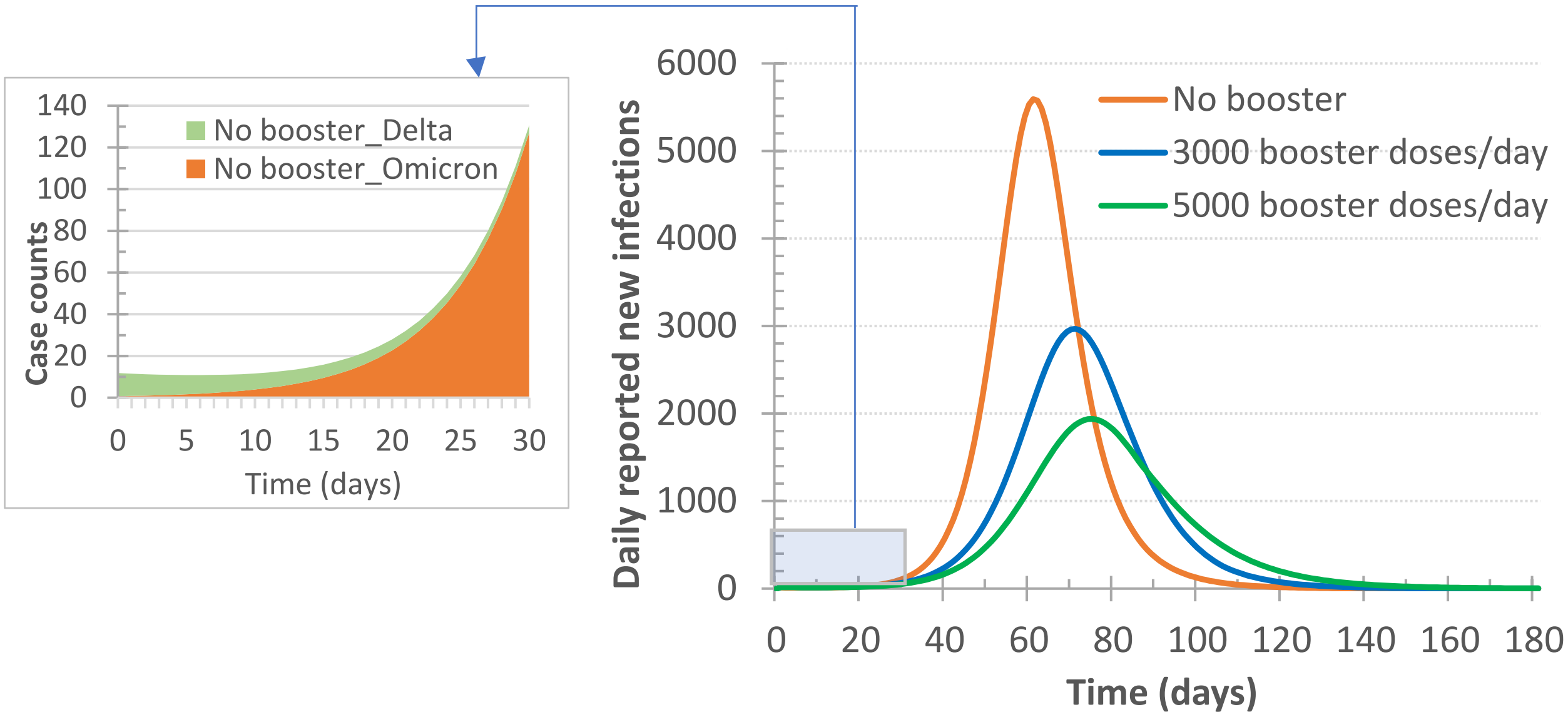

Figure 6: Impact of booster doses on COVID-19 transmission with two variants

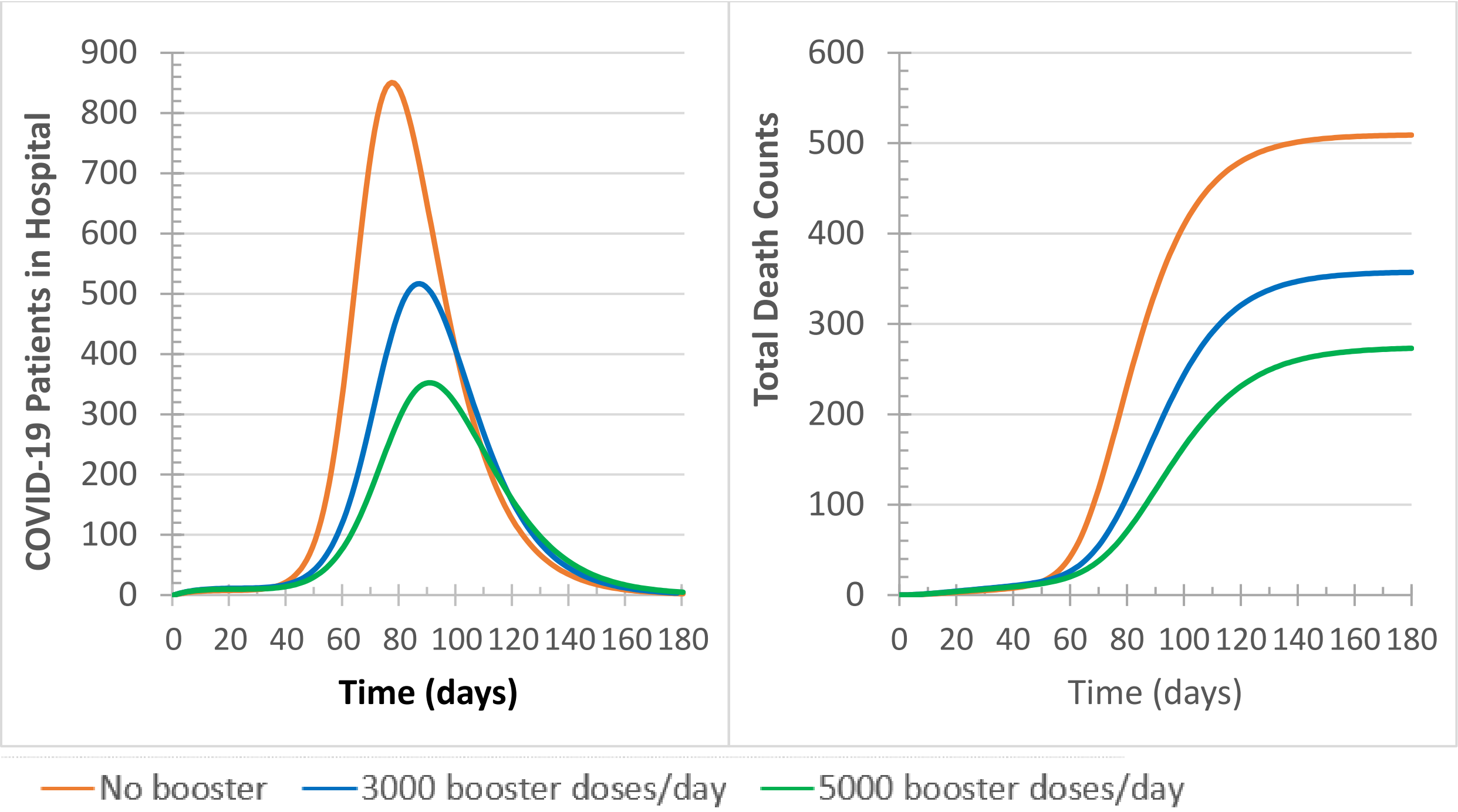

Figure 7: Impact of booster on severity of COVID-19 infections with two variants

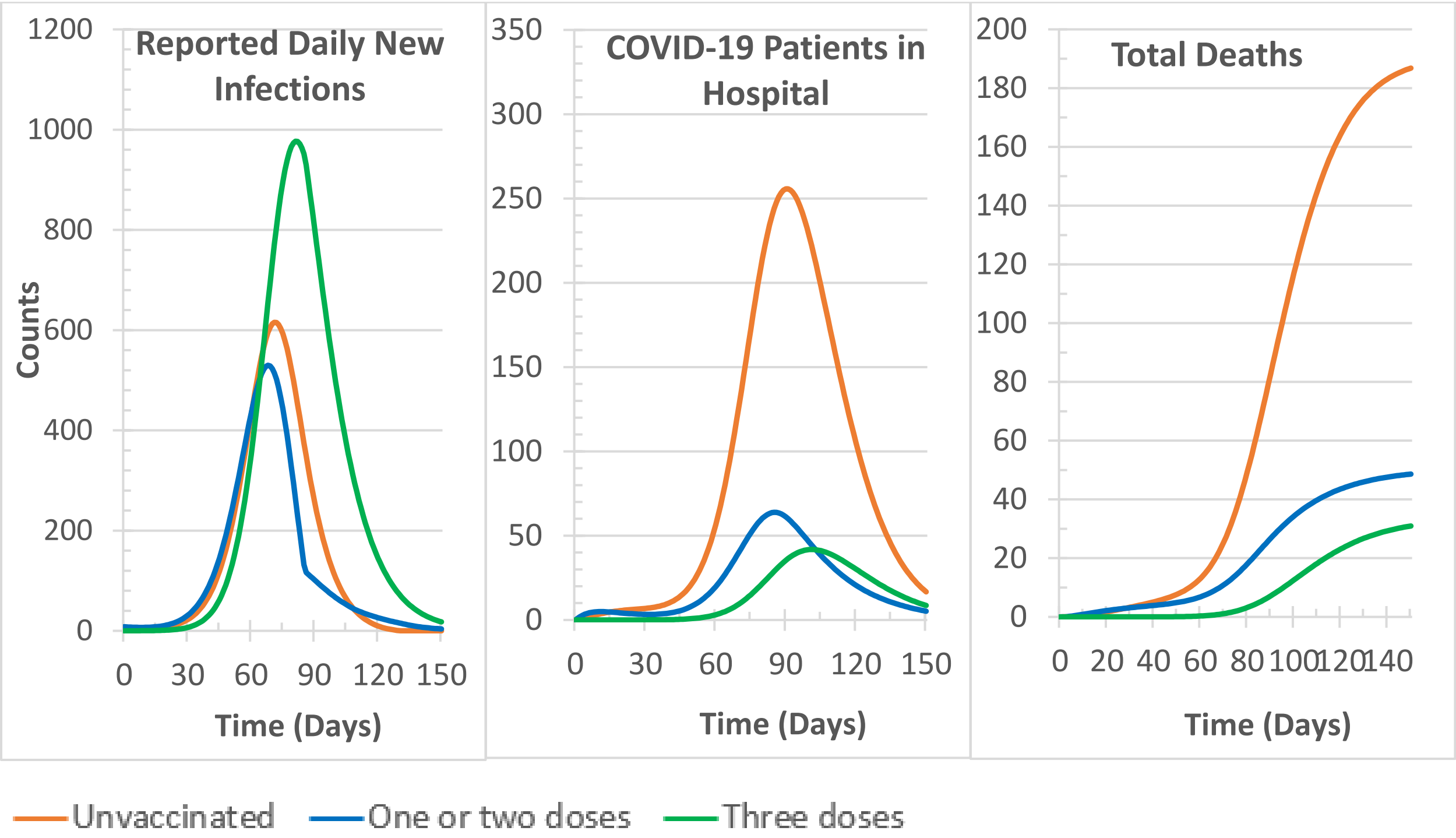

Figure 8: COVID-19 infections, hospitalizations and deaths by vaccine status with 5000 booster doses administered per day

Table 3: Counts of infections, hospitalizations and deaths over the simulation period (180 days) and percentage decrease compared to the “No-booster” scenario

| Groups | Total infections* | % decrease in total infections | Highest hospitalizations on a single day | % decrease in hospitalization peak | Total hospitalizations | % decrease in total hospitalizations | Total deaths | % decrease in total deaths |
|---|---|---|---|---|---|---|---|---|
| No booster | 558,841 | - | 851 | - | 2558 | | 509 | |
| 3000 booster doses/day | 411,500 | 26% | 517 | 39% | 1766 | 31% | 357 | 30% |
| 5000 booster doses/day | 328,533 | 41% | 352 | 59% | 1354 | 47% | 273 | 46% |

Table 4: Counts and proportion of infections, hospitalizations and deaths over the simulation period (180 days) by vaccination status assuming 5000 booster doses administered per day

| | Counts | | | | Proportions | | |
|---|---|---|---|---|---|---|---|
| | Unvaccinated | 1 or 2 doses | 3 doses | Total | Unvaccinated | 1 or 2 doses | 3 doses |
| Total infections | 90,288 | 74,919 | 163,327 | 328,533 | 27% | 23% | 50% |
| Hospitalization Peak | 256 | 64 | 42 | 362 | 71% | 18% | 12% |
| Total hospitalizations | 946 | 241 | 167 | 1354 | 70% | 18% | 12% |
| Total deaths | 190 | 50 | 33 | 273 | 70% | 18% | 12% |

## 5. CONCLUSIONS AND FUTURE INVESTIGATIONS

A new, easy-to-use epidemiological-modelling desktop app was developed based on a multi-compartment deterministic epidemiological model. The app is able to model different levels of vaccine-induced immunity, as well as the developing and waning of immunity with time after vaccination. The functionality of the app is demonstrated by using it to simulate the effects of specific factors on COVID-19 transmission. Simulation results yield several conclusions:

- For the Delta variant, herd immunity is not achievable through vaccination only. To maintain a reproduction number below 1, public health measures need to be in place until natural immunity achieved through infection with the virus, along with immunity through vaccination, brings the overall immunity to the level necessary for herd immunity.
- Waning vaccine-induced immunity prolongs the time public health measures need to stay in place and the time necessary to reach herd immunity through additional infections.

- The Omicron variant quickly outcompetes the Delta variant within two weeks and the number of daily new cases is projected to start decreasing after two to three months, depending on the administration of booster doses.
- Booster doses have an important contribution to mitigating the effects of waning immunity and immune evasion by reducing COVID-19 infections, hospitalizations and deaths.

Limitations of the current model include the assumption of full and permanent immunity after infection and the assumption that infection with one variant will offer immunity against all other variants.

The model is currently being extended to include options to model re-infection with either the same or a different variant, as well as stratum-specific number of contacts per day. The inclusion of these new features will allow more realistic simulations, including the study of annual, possibly seasonal, epidemics under endemic conditions.

While COVID-19 provided the impetus for this work, the developed model and desktop app are flexible enough to be applicable to other communicable diseases being monitored by PHUs. Thus, it is expected that IDSIM will be a welcome addition to the tools in current use by epidemiologists in PHUs.

## 6. APP AVAILABILITY

The app can be downloaded from www.idsim.ca.

## 7. ACKNOWLEDGMENT

The reported work was supported in part by an Alliance COVID-19 grant from the Canadian Natural Sciences and Engineering Research Council (NSERC).

## APPENDIX: MODEL EQUATIONS

### A.1 Notations

The equations in this appendix use the following notations.

***General parameters, quantities and identifiers***

$N_s$ Number of stratifications

$k$ Stratification index $(k = 1, 2 ... N_s)$

$n_k$ Number of strata in stratification $k$

$s_k$ stratum index for stratification $k$ $(s_k = 1, 2 ... n_k)$

$g$ Combined-stratum index $(g = \{s_1, s_2 ... s_{N_s}\})$

$n_s$ Total number of combined strata $(n_s = \prod_{k=1}^{N_s} n_k)$

$C$ Compartment identifier. The compartment identifiers are described in section 2.1 of the main paper. For example, $C$=$EQ$ denotes the "Exposed Quarantined" compartment.

$N_{g,i,\text{v}}^{C}(t)$ Number of individuals in compartment $C$ belonging to combined stratum $g$, with immunization level $i$, affected by variant v, at time $t$. For example, $N_{g,i,\text{v}}^{EQ}(t)$ denotes the number of exposed quarantined individuals.

$N$ Total population

$\chi$ Probability of transmission with contact (with an infectious individual)

$\Phi$ Contact rate [number of contacts (with other individuals) a (susceptible) individual has per unit time (day)]

***Stratification parameters***

$\alpha_{k,s_k}^{sus}$ Susceptibility modulator for stratum $s_k$ of stratification $k$

$\alpha_{g}^{sus}$ Susceptibility modulator for combined stratum $g$ $(\alpha_{g}^{sus} = \prod_{k=1}^{N_s} \alpha_{k,s_k}^{sus})$

$\alpha^{sev}_{k,s_k}$ Severity modulator for stratum $s_k$ of stratification $k$

$\alpha^{sev}_{g}$ Severity modulator for combined stratum $g$ $(\alpha^{sev}_{g} = \prod_{k=1}^{N_s} \alpha^{sev}_{k,s_k})$

***Variant parameters***

$T^{lat}_{\mathrm{v}}$ Latency time (since exposure) for variant v

$T^{inc}_{\mathrm{v}}$ Incubation time (since exposure) for variant v

$T^{hos}_{\mathrm{v}}$ Time to hospitalization for severe cases (since exposure) for variant v

$T^{rec-ns}_{\mathrm{v}}$ Time to recovery for non-severe cases (since exposure) for variant v

$T^{hos-rec}_{\mathrm{v}}$ Time to recovery after hospitalization (for severe cases that recover)

$T^{hos-dec}_{\mathrm{v}}$ Time to death after hospitalization (for severe cases that do not recover)

$\chi_{\mathrm{v}}$ Probability of transmission with contact

$\gamma^{sym}_{\mathrm{v}}$ Fraction of infectious individuals that are symptomatic

$\gamma^{sev}_{\mathrm{v}}$ Fraction of infectious symptomatic individuals that have severe symptoms

$\gamma^{hos-rec}_{\mathrm{v}}$ Fraction of hospitalized individuals that recover

***Immunization parameters***

$\theta^{tra}_{i,\mathrm{v}}$ Transmissibility factor for variant v and immunization level *i*

$\theta^{sus}_{i,\mathrm{v}}$ Susceptibility factor for variant v and immunization level *i*

$\theta^{sev}_{i,\mathrm{v}}$ Severity factor for variant v and immunization level *i*

$R^{vac}_{i}$ Persons with immunity level *i* vaccinated per unit time (day)

$T^{vac}_{i}$ Time (days) spent in immunity level *i* before advancing to immunity level *i+1*

NOTE: For immunization levels, *i*, for which progression to level *i+1* happens through vaccination, $R^{vac}_{i} \neq 0$ and $\frac{1}{T^{vac}_{i}} = 0$. For immunization levels, *i*, for which progression to level *i+1* happens through the simple passage of time (such as in the case of developing protection

after vaccination or in the case of protection waning), $R_i^{vac} = 0$ and $\frac{1}{T_i^{vac}} \neq 0$. In short, either $R_i^{vac}$ or $T_i^{vac}$, but not both, apply to any immunity level *i*, and $\frac{R_i^{vac}}{T_i^{vac}} = 0$.

***Parameters for public health measures***

$\gamma^{q}$ Fraction of exposed individuals that are successfully quarantined

$\gamma^{ti}$ Fraction of infectious individuals that are tested and successfully isolated

$\sigma^{phm}$ Coefficient for additional, unspecified, public health measures

## A.2 Force of infection

The force (risk) of infection is a susceptible individual's probability of exposure per unit time. The force of infection is denoted by $\lambda_{g,i,\text{v}}(t)$ and has the following expression:

$$\lambda_{g,i,\text{v}}(t) = \sigma^{phm} \alpha_g^{sus} \theta_{i,\text{v}}^{sus} \sum_{i'} \theta_{i',\text{v}}^{tra} \chi_{\text{v}} \Phi \frac{1}{N - \sum_{g'',i'',\text{v}''} N_{g'',i'',\text{v}''}^{D}(t)} \sum_{g'} \sum_{\substack{C \in \text{infectious in} \\ \text{the community}}} N_{g',i',\text{v}}^{C}(t) \tag{1}$$

For a small number of deaths, $\sum_{g',i'',\text{v}'} N_{g',i'',\text{v}'}^{D}(t) \ll N$, the force of infection can be approximated by:

$$\lambda_{g,i,\text{v}}(t) \simeq \sigma^{phm} \alpha_g^{sus} \theta_{i,\text{v}}^{sus} \sum_{i'} \theta_{i',\text{v}}^{tra} \chi_{\text{v}} \Phi \frac{1}{N} \sum_{g'} \sum_{\substack{C \in \text{infectious in} \\ \text{the commuity}}} N_{g',i',\text{v}}^{C}(t) \tag{2}$$

## A.3 Vaccination

Only individuals who are susceptible or otherwise in the community (neither isolated nor quarantined) and not symptomatic are vaccinated under normal circumstances. Consequently, the number of "vaccinable" persons with immunization level *i*, at any given time is:

$$\begin{aligned} N_i^{vac}(t) = {} & \sum_{g} N_{g,i}^{S}(t) + \sum_{g} \sum_{\text{v}} N_{g,i,\text{v}}^{EC}(t) + \sum_{g} \sum_{\text{v}} N_{g,i,\text{v}}^{ICA}(t) \\ & \sum_{g} \sum_{\text{v}} N_{g,i,\text{v}}^{ICPM}(t) + \sum_{g} \sum_{\text{v}} N_{g,i,\text{v}}^{ICPS}(t) \end{aligned} \tag{3}$$

If exposed and infectious individuals are much fewer than the susceptible ones, it can be assumed that only susceptible individuals are being vaccinated and then the number of "vaccinable" individuals can be approximated as:

$$N_i^{vac}(t) \simeq \sum_g N_{g,i}^S(t) \tag{4}$$

This work assumes that only susceptible individuals are being vaccinated.

With the above notations, the balance equations for each compartment are written as below.

### A.4 Balance equations

$$\begin{aligned}\frac{d}{dt}N_{g,i}^S(t) = -\left(\sum_{\mathrm{v}} \lambda_{g,i,\mathrm{v}}(t)\right) \times N_{g,i}^S(t) + \\ R_{i-1}^{vac}\frac{N_{g,i-1}^S(t)}{N_{i-1}^{vac}(t)} + \frac{1}{T_{i-1}^{vac}}N_{g,i-1}^S(t) - R_i^{vac}\frac{N_{g,i}^S(t)}{N_i^{vac}(t)} - \frac{1}{T_i^{vac}}N_{g,i}^S(t)\end{aligned} \tag{5}$$

The fourth and fifth terms on the right represent the rate at which persons with current immunization level *i* move to immunization level *i+1*. As explained in the note for the immunization parameters, only one of the two terms is nonzero. Similarly, terms two and three on the right represent the rate at which persons with current immunization level *i-1* move to immunization level *i*, and only one of them is nonzero.

$$\frac{d}{dt}N_{g,i,\mathrm{v}}^{EC}(t) = \left(1-\gamma^q\right)\lambda_{g,i,\mathrm{v}}(t) \times N_{g,i,\mathrm{v}}^S(t) + \frac{1}{T_{i-1}^{vac}}N_{g,i-1,\mathrm{v}}^{EC}(t) - \frac{1}{T_i^{vac}}N_{g,i,\mathrm{v}}^{EC}(t) \tag{6}$$

$$\frac{d}{dt}N_{g,i,\mathrm{v}}^{EQ}(t) = \gamma^q\lambda_{g,i,\mathrm{v}}(t) \times N_{g,i,\mathrm{v}}^S(t) + \frac{1}{T_{i-1}^{vac}}N_{g,i-1,\mathrm{v}}^{EQ}(t) - \frac{1}{T_i^{vac}}N_{g,i,\mathrm{v}}^{EQ}(t) \tag{7}$$

$$\begin{aligned}\frac{d}{dt}N_{g,i,\mathrm{v}}^{ICA}(t) = \left(1-\gamma_{\mathrm{v}}^{sym}\right)\left(1-\gamma^{ti}\right)\frac{1}{T_{\mathrm{v}}^{lat}} \times N_{g,i,\mathrm{v}}^{EC}(t) - \frac{1}{T_{\mathrm{v}}^{rec-ns} - T_{\mathrm{v}}^{lat}} \times N_{g,i,\mathrm{v}}^{ICA}(t) + \\ \frac{1}{T_{i-1}^{vac}}N_{g,i-1,\mathrm{v}}^{ICA}(t) - \frac{1}{T_i^{vac}}N_{g,i,\mathrm{v}}^{ICA}(t)\end{aligned} \tag{8}$$

$$\begin{aligned}\frac{d}{dt}N_{g,i,\mathrm{v}}^{ICPM}(t) = \left(1-\alpha_g^{sev}\gamma_{\mathrm{v}}^{sev}\right)\gamma_{\mathrm{v}}^{sym}\left(1-\gamma^{ti}\right)\frac{1}{T_{\mathrm{v}}^{lat}} \times N_{g,i,\mathrm{v}}^{EC}(t) - \frac{1}{T_{\mathrm{v}}^{inc} - T_{\mathrm{v}}^{lat}} \times N_{g,i,\mathrm{v}}^{ICPM}(t) + \\ \frac{1}{T_{i-1}^{vac}}N_{g,i-1,\mathrm{v}}^{ICPM}(t) - \frac{1}{T_i^{vac}}N_{g,i,\mathrm{v}}^{ICPM}(t)\end{aligned} \tag{9}$$

$$\frac{d}{dt}N_{g,i,v}^{ICPS}(t)=\alpha_g^{sev}\gamma_v^{sev}\gamma_v^{sym}\left(1-\gamma^{ti}\right)\frac{1}{T_v^{lat}}\times N_{g,i,v}^{EC}(t)-\frac{1}{T_v^{inc}-T_v^{lat}}\times N_{g,i,v}^{ICPS}(t)+\frac{1}{T_{i-1}^{vac}}N_{g,i-1,v}^{ICPS}(t)-\frac{1}{T_i^{vac}}N_{g,i,v}^{ICPS}(t) \tag{10}$$

$$\frac{d}{dt}N_{g,i,v}^{IIA}(t)=\left(1-\gamma_v^{sym}\right)\gamma^{ti}\frac{1}{T_v^{lat}}\times N_{g,i,v}^{EC}(t)-\frac{1}{T_v^{rec-ns}-T_v^{lat}}\times N_{g,i,v}^{IIA}(t)+\frac{1}{T_{i-1}^{vac}}N_{g,i-1,v}^{IIA}(t)-\frac{1}{T_i^{vac}}N_{g,i,v}^{IIA}(t) \tag{11}$$

$$\frac{d}{dt}N_{g,i,v}^{IIPM}(t)=\left(1-\alpha_g^{sev}\gamma_v^{sev}\right)\gamma_v^{sym}\gamma^{ti}\frac{1}{T_v^{lat}}\times N_{g,i,v}^{EC}(t)-\frac{1}{T_v^{inc}-T_v^{lat}}\times N_{g,i,v}^{IIPM}(t)+\frac{1}{T_{i-1}^{vac}}N_{g,i-1,v}^{IIPM}(t)-\frac{1}{T_i^{vac}}N_{g,i,v}^{IIPM}(t) \tag{12}$$

$$\frac{d}{dt}N_{g,i,v}^{IIPS}(t)=\alpha_g^{sev}\gamma_v^{sev}\gamma_v^{sym}\gamma^{ti}\frac{1}{T_v^{lat}}\times N_{g,i,v}^{EC}(t)-\frac{1}{T_v^{inc}-T_v^{lat}}\times N_{g,i,v}^{IIPS}(t)+\frac{1}{T_{i-1}^{vac}}N_{g,i-1,v}^{IIPs}(t)-\frac{1}{T_i^{vac}}N_{g,i,v}^{IIPS}(t) \tag{13}$$

$$\frac{d}{dt}N_{g,i,v}^{IQA}(t)=\left(1-\gamma_v^{sym}\right)\frac{1}{T_v^{lat}}\times N_{g,i,v}^{EQ}(t)-\frac{1}{T_v^{rec-ns}-T_v^{lat}}\times N_{g,i,v}^{IQA}(t)+\frac{1}{T_{i-1}^{vac}}N_{g,i-1,v}^{IQA}(t)-\frac{1}{T_i^{vac}}N_{g,i,v}^{IQA}(t) \tag{14}$$

$$\frac{d}{dt}N_{g,i,v}^{IQPM}(t)=\left(1-\alpha_g^{sev}\gamma_v^{sev}\right)\gamma_v^{sym}\frac{1}{T_v^{lat}}\times N_{g,i,v}^{EQ}(t)-\frac{1}{T_v^{inc}-T_v^{lat}}\times N_{g,i,v}^{IQPM}(t)+\frac{1}{T_{i-1}^{vac}}N_{g,i-1,v}^{IQPM}(t)-\frac{1}{T_i^{vac}}N_{g,i,v}^{IQPM}(t) \tag{15}$$

$$\frac{d}{dt}N_{g,i,v}^{IQPS}(t)=\alpha_g^{sev}\gamma_v^{sev}\gamma_v^{sym}\frac{1}{T_v^{lat}}\times N_{g,i,v}^{EQ}(t)-\frac{1}{T_v^{inc}-T_v^{lat}}\times N_{g,i,v}^{IQPS}(t)+\frac{1}{T_{i-1}^{vac}}N_{g,i-1}^{IQPS}(t)-\frac{1}{T_i^{vac}}N_{g,i}^{IQPS}(t) \tag{16}$$

$$\frac{d}{dt}N_{g,i,v}^{ICSM}(t)=\frac{1}{T_v^{inc}-T_v^{lat}}\times N_{g,i,v}^{ICPM}(t)-\frac{1}{T_v^{rec-ns}-T_v^{inc}}\times N_{g,i,v}^{ICSM}(t)+\frac{1}{T_{i-1}^{vac}}N_{g,i-1,v}^{ICSM}(t)-\frac{1}{T_i^{vac}}N_{g,i,v}^{ICSM}(t) \tag{17}$$

$$\frac{d}{dt}N_{g,i,v}^{ICSS}(t)=\frac{1}{T_v^{inc}-T_v^{lat}}\times N_{g,i,v}^{ICPS}(t)-\frac{1}{T_v^{hos}-T_v^{inc}}\times N_{g,i,v}^{ICSS}(t)+ \\ \frac{1}{T_{i-1}^{vac}}N_{g,i-1,v}^{ICSS}(t)-\frac{1}{T_i^{vac}}N_{g,i,v}^{ICSS}(t) \tag{18}$$

$$\frac{d}{dt}N_{g,i,v}^{IISM}(t)=\frac{1}{T_v^{inc}-T_v^{lat}}\times N_{g,i,v}^{IIPM}(t)-\frac{1}{T_v^{rec-ns}-T_v^{inc}}\times N_{g,i,v}^{IISM}(t)+ \\ \frac{1}{T_{i-1}^{vac}}N_{g,i-1,v}^{IISM}(t)-\frac{1}{T_i^{vac}}N_{g,i,v}^{IISM}(t) \tag{19}$$

$$\frac{d}{dt}N_{g,i,v}^{IISS}(t)=\frac{1}{T_v^{inc}-T_v^{lat}}\times N_{g,i,v}^{IIPS}(t)-\frac{1}{T_v^{hos}-T_v^{inc}}\times N_{g,i,v}^{IISS}(t)+ \\ \frac{1}{T_{i-1}^{vac}}N_{g,i-1,v}^{IISS}(t)-\frac{1}{T_i^{vac}}N_{g,i,v}^{IISS}(t) \tag{20}$$

$$\frac{d}{dt}N_{g,i,v}^{IQSM}(t)=\frac{1}{T_v^{inc}-T_v^{lat}}\times N_{g,i,v}^{IQPM}(t)-\frac{1}{T_v^{rec-ns}-T_v^{inc}}\times N_{g,i,v}^{IQSM}(t)+ \\ \frac{1}{T_{i-1}^{vac}}N_{g,i-1,v}^{IQSM}(t)-\frac{1}{T_i^{vac}}N_{g,i,v}^{IQSM}(t) \tag{21}$$

$$\frac{d}{dt}N_{g,i,v}^{IQSS}(t)=\frac{1}{T_v^{inc}-T_v^{lat}}\times N_{g,i,v}^{IQPS}(t)-\frac{1}{T_v^{hos}-T_v^{inc}}\times N_{g,i,v}^{IQSS}(t)+ \\ \frac{1}{T_{i-1}^{vac}}N_{g,i-1,v}^{IQSS}(t)-\frac{1}{T_i^{vac}}N_{g,i,v}^{IQSS}(t) \tag{22}$$

$$\frac{d}{dt}N_{g,i,v}^{HR}(t)=\gamma_v^{hos-rec}\times\frac{1}{T_v^{hos}-T_v^{inc}}\times\left[N_{g,i,v}^{ICSS}(t)+N_{g,i,v}^{IISS}(t)+N_{g,i,v}^{IQSS}(t)\right]- \\ \frac{1}{T_v^{hos-rec}}\times N_{g,i,v}^{HR}(t)+\frac{1}{T_{i-1}^{vac}}N_{g,i-1,v}^{HR}(t)-\frac{1}{T_i^{vac}}N_{g,i,v}^{HR}(t) \tag{23}$$

$$\frac{d}{dt}N_{g,i,v}^{HD}(t)=\left(1-\gamma_v^{hos-rec}\right)\times\frac{1}{T_v^{hos}-T_v^{inc}}\times\left[N_{g,i,v}^{ICSS}(t)+N_{g,i,v}^{IISS}(t)+N_{g,i,v}^{IQSS}(t)\right]- \\ \frac{1}{T_v^{hos-dec}}\times N_{g,i,v}^{HD}(t)+\frac{1}{T_{i-1}^{vac}}N_{g,i-1,v}^{HD}(t)-\frac{1}{T_i^{vac}}N_{g,i,v}^{HD}(t) \tag{24}$$

$$\frac{d}{dt}N_{g,i,\text{v}}^{R}(t)=\frac{1}{T_{\text{v}}^{rec-ns}-T_{\text{v}}^{lat}}\times N_{g,i,\text{v}}^{ICA}(t)+\frac{1}{T_{\text{v}}^{rec-ns}-T_{\text{v}}^{inc}}\times N_{g,i,\text{v}}^{ICSM}(t)+$$
$$\frac{1}{T_{\text{v}}^{rec-ns}-T_{\text{v}}^{lat}}\times N_{g,i,\text{v}}^{IIA}(t)+\frac{1}{T_{\text{v}}^{rec-ns}-T_{\text{v}}^{inc}}\times N_{g,i,\text{v}}^{IISM}(t)+$$
$$\frac{1}{T_{\text{v}}^{rec-ns}-T_{\text{v}}^{lat}}\times N_{g,i,\text{v}}^{IQA}(t)+\frac{1}{T_{\text{v}}^{rec-ns}-T_{\text{v}}^{inc}}\times N_{g,i,\text{v}}^{IQSM}(t)+$$
$$\frac{1}{T_{\text{v}}^{hos-rec}}\times N_{g,i,\text{v}}^{HR}(t)+\frac{1}{T_{i-1}^{vac}}N_{g,i-1,\text{v}}^{R}(t)-\frac{1}{T_{i}^{vac}}N_{g,i,\text{v}}^{R}(t) \quad (25)$$

$$\frac{d}{dt}N_{g,i,\text{v}}^{D}(t)=\frac{1}{T_{\text{v}}^{hos-dec}}\times N_{g,i,\text{v}}^{HD}(t) \quad (26)$$

In balance equations (6) to (25), corresponding to any compartment *C* other than *S*, the two terms on the right-hand side of type $\frac{1}{T_{i-1}^{vac}}N_{g,i-1,\text{v}}^{C}(t)$ and $\frac{1}{T_{i}^{vac}}N_{g,i,\text{v}}^{C}(t)$ represent, respectively, the rate at which persons with current immunization level *i-1* move to immunization level *i* and the rate at which persons with current immunization level *i* move to immunization level *i+1*, through the passage of time. Depending on the desired type of simulation, one or both terms can be zero (See also previous note for the immunization parameters.).